# The Uniformization Process of the Fast Congestion Notification (FN)

Mohammed M. Kadhum
InterNetWorks Research Group
College of Arts and Sciences
Universiti Utara Malaysia
06010 UUM Sintok, Malaysia
kadhum@uum.edu.my

Suhaidi Hassan
InterNetWorks Research Group
College of Arts and Sciences
Universiti Utara Malaysia
06010 UUM Sintok, Malaysia
suhaidi@uum.edu.my

*Abstract*—**Fast Congestion Notification (FN) is one of the proactive queue management mechanisms that practices congestion avoidance to help avoid the beginning of congestion by marking/dropping packets before the router's queue gets full; and exercises congestion control, when congestion avoidance fails, by increasing the rate of packet marking/dropping. Technically, FN avoids the queue overflows by controlling the instantaneous queue size below the optimal queue size, and control congestion by keeping the average arrival rate close to the outgoing link capacity. Upon arrival of each packet, FN uses the instantaneous queue size and the average arrival rate to calculate the packet marking/dropping probability. FN marks/drops packets at fairly regular intervals to avoid long intermarking intervals and clustered packet marks/drops. Too many marked/dropped packets close together can cause global synchronization, and also too long packet intermarking times between marked/dropped packets can cause large queue sizes and congestion. This paper shows how FN controls the queue size, avoids congestion, and reduces global synchronization by uniformizing marked/dropped packet intervals.**

## I. INTRODUCTION

Internet gateways' queues are used to accommodate incoming packet and to allow the gateway enough time for packet transmission. When the arriving packet rate is higher than the gateway's outgoing link capacity, the queue size will increase, until the gateway buffer becomes full. When the buffer is full, the newly arriving packet will be dropped.

In the current Internet, the TCP transport protocol detects congestion only after a packet has been marked/dropped at the gateway. However, it would clearly be undesirable to have large queues that were full much of the time; this would significantly increase the average delay in the network. Hence, with increasingly high-speed networks, it is important to have mechanisms that keep throughput high but average queue sizes low [1].

Active queue management (AQM) mechanisms mark/drop packets before the gateway's buffer is full. These mechanisms operate by maintaining one or more mark/drop probabilities, and probabilistically dropping/marking packets even when the queue is short.

## II. ACTIVE QUEUE MANAGEMENT (AQM)

Active queue management policies, such as Random Early Detection (RED), are expected to eliminate global synchronization that introduced by reactive queue management policies and improve Quality of Service (QoS) of the networks. The promised advantages of AQM are increase in throughput, reduce the delay, high link utilizations, and avoid lock-out. AQM provides preventive measures to manage the router queue to overcome the problems associated with passive queue management policies. AQM has the following attributes:

- Performing a preventive random packet mark/drop before the queue is full.
- The probability of the preventive packet mark/drop is proportional to congestion levels.

Preventive packet mark/drop provides implicit feedback method to notify the traffic senders of the congestion onset [2]. As a reaction, senders reduce their transmission rate to moderate the congestion level. Arriving packets from the senders are marked/dropped randomly, which prevents senders from backing off at the same time and thereby eliminate global synchronization [2].

Different packet marking/dropping strategies have different impacts on the gateway performance, including packet delays, number of dropped packets, and link utilizations. Generally, with a given AQM scheme, if a gateway drops packets more aggressively, less packets will be admitted and go through the gateway, hence the outgoing link's utilization may be lower; but in return, the admitted packets will experience smaller delays. Conversely, if under an AQM scheme which drops packets less aggressively, the admitted packets may be queued up at the gateway, hence the admitted packets will experience larger delays. But in this case the outgoing link's utilization may be higher, since more packets are admitted and transmitted by the gateway [3].

### A. Random Early Detection (RED)

RED [1] is one of AQM mechanisms that requires the user to specify five parameters: the maximum buffer size or queue limit (*QL*), the minimum (*min$_{th}$*) and maximum (*max$_{th}$*) thresholds of the "RED region", the maximum dropping probability (*max$_p$*), and the weight factor used to calculate the average queue size (*w$_q$*). *QL* can be defined in terms of packets or bytes. A RED gateway uses early packet dropping in an attempt to control the congestion level, limit queuing delays, and avoid buffer overflows. Early packet dropping starts when the average queue size exceeds *min$_{th}$*. RED was specifically designed to use the average queue size (*avg*), instead of the current queue size, as a measure of incipient congestion, because the latter proves to be rather intolerant of packet bursts. If the average queue size does not exceed *min$_{th}$*, a RED gateway will not drop any packet. *avg* is calculated as an exponentially weighted moving average using the following formula:

$$avg_i = (1 - w_q) \times avg_{i-1} + w_q \times q \quad (1)$$

where the weight *w$_q$* is commonly set to 0.002, and *q* is the instantaneous queue size. This weighted moving average captures the notion of long-lived congestion better than the instantaneous queue size [4]. Had the instantaneous queue size been used as the metric to determine whether the router is congested, short-lived traffic spikes would lead to early packet drops. So a rather underutilized router that receives a burst of packets can be deemed "congested" if one uses the instantaneous queue size. The average queue size, on the other hand, acts as a low pass filter that allows spikes to go through the router without forcing any packet drops (unless, of course, the burst is larger than the queue limit). The user can configure *w$_q$* and *min$_{th}$* so that a RED router does not allow short-lived congestion to continue uninterrupted for more than a predetermined amount of time. This functionality allows RED to maintain high throughput and keep per-packet delays low.

RED uses randomization to decide which packet to drop and, consequently, which connection will be notified to slow down. This is accomplished using a probability *p$_a$*, which is calculated according to the following formulae:

$$p_b = max_p \times (avg - min_{th}) / (max_{th} - min_{th}) \quad (2)$$

and

$$p_a = p_b / (1 - count \times p_b) \quad (3)$$

where maxp is a user-defined parameter, usually set to 2% or 10%, and *count* is the number of packets since the last packet mark/drop. *count* is used so that consecutive marks/drops are spaced out over time. Notice that *p$_b$* varies linearly between 0 and *max$_p$*, while *p$_a$*, i.e., the actual packet marking/dropping probability increases with *count* [5]. Originally, *max$_{th}$* was defined as the upper threshold; when the average queue size exceeds this limit, all packets have to be dropped (Figure 1(a)). Later on, a gentle version of RED [6] was proposed as a modification to the dropping algorithm, under which packets are dropped with a linearly increasing probability until *avg* exceeds 2×*max$_{th}$*; after that all packets are dropped (Figure 1(b)). Although *max$_{th}$* can be set to any value, a rule of thumb is to set it to three times *min$_{th}$*, and less than *QL* [4].

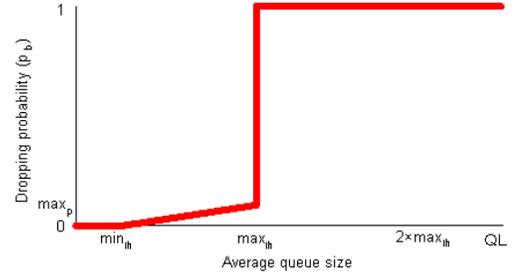

(a) RED

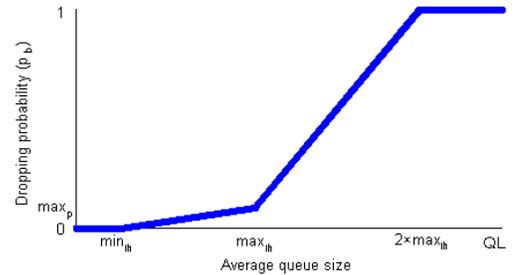

(b) RED with the "gentle option"

Figure 1. The packet dropping probability (*p$_b$*) in RED as a function of the average queue size (*max$_p$* = 10%)

By marking/dropping packets before the buffer overflows, RED attempts to notify some connections of incipient congestion. The responsive ones will limit their sending rates and eventually the network load will decrease. The unresponsive connections will not slow down, but will continue at the same pace or even increase their sending rates. In this case, the unresponsive flows will have more packets reaching the router, effectively providing more candidates for dropping than responsive ones.

### B. Fast Congestion Notification (FN)

The Fast Congestion Notification (FN) [7] queue management algorithm randomly marks (if ECN) / drops (if non-ECN) the arriving packets before the buffer overflows, to effectively control the:

- Instantaneous queue length below the optimal queue length to reduce the queuing delay and avoid the buffer overflows.

- Average traffic arrival rate of the queue in the proximity of the departing link capacity to enable the congestion and queue length control.

FN integrates the instantaneous queue length and the average arrival rate of queue to compute the mark/drop probability of the packet upon each arriving packet. The use of the instantaneous queue length in conjunction with the average queue speed (average arrival rate) can provide superior control decision criteria for an active queue management scheme [8].

The FN linear mark/drop probability function [9] is derived based on the assumption that the arrival traffic process remains unchanged over the control time constant period of length ($T$) seconds. In other words, it is supposed that immediately following the packet's arrival, the traffic continues to arrive at the fixed rate of ($R$) bits/sec, the estimated average arrival rate to the buffer computed upon the packet's arrival, for the period of the control time constant. The buffer has a capacity of ($C$) bits and is served by an outgoing link at a fixed rate of ($\mu$) bits/sec. The packet mark/drop probability ($P$), is computed for, and applied to, every incoming packet, based on the above assumptions, with the goal of driving the instantaneous (current) queue length ($Q_{cur}$) to some desired optimal level ($Q_{opt}$) over the control time constant period ($T$). These are shown in figure 2. The FN mark/drop probability, $P$, is calculated by

$$P^{(i)} = \frac{((R_i - \mu).T) - (Q_{opt} - Q_{cur})}{R_i.T} \quad (4)$$

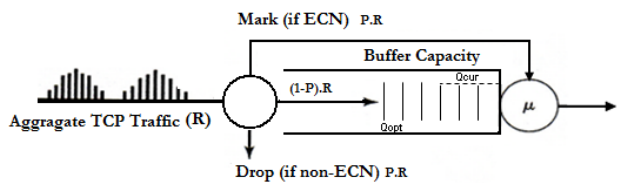

Figure 2. FN Gateway Buffer

### III. UNIFORMIZATION OF PACKET MARKS/DROPS

An attractive property of RED resulting from using the *count* variable is that the number of accepted packets between two packet marks/drops is uniformly distributed [1]. By having a uniform distribution, packet marks/drops are not clustered, avoiding again possible synchronization of TCP sources. Although quantitative benefits of having a uniform distribution were not, at the best of our knowledge, reported in the literature it is commonly admitted that having light-tailed distributions (such as the uniform distribution) gives better performance in terms of efficiency and fairness [5].

Same as RED, FN marks/drops packets at fairly regular intervals. FN uniformization technique enforces packet marks/drops at evenly spaced intervals to avoid long periods of time, where no packets are marked or dropped and clustered packet marks/drops, under the steady-state conditions at gateway. Very long packet marking/dropping times can contribute to large queue sizes and congestion. Multiple successive packet marks/drops can result in global synchronization problem.

To compute the initial marking/dropping probability, FN uses the average traffic arrival rate and the instantaneous queue size by

$$P_{ini} = \frac{((R - \mu).T) + (Q_{cur} - Q_{opt})}{R.T} \quad (5)$$

The initial marking/dropping probability is used along with the number of accepted packets between two packet marks/drops (*count*) by the uniformization function to calculate the final packet marking/dropping probability as follows:

$$P_{fin} = \begin{cases} \dfrac{P_{ini}}{2-count.P_{ini}} & count.P_{ini} < 2 \\ 1 & otherwise \end{cases} \quad (6)$$

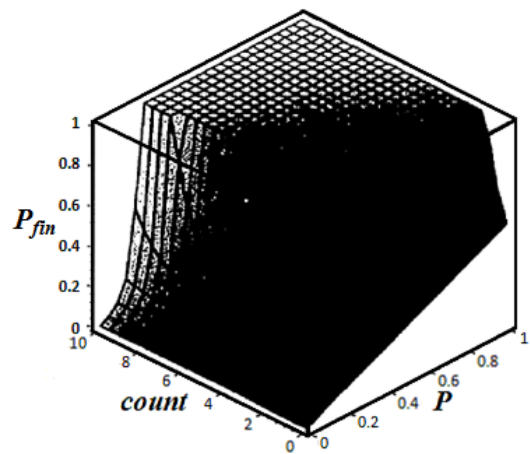

Figure 1. FN Uniformization Function - $P_{fin} = \dfrac{P_{ini}}{2 - count.P_{ini}}$

Figure 3 shows that the FN uniformization function increases the value of the initial marking/dropping probability proportional to the number of the accepted packets (*count*) since the last marking/dropping. When *count* increases, the final marking/dropping probability $P_{fin}$ will rise until it finally reaches 1. This means that even if the conditions at the gateway are such that $P$ remains comparatively constant, the uniformization technique directs the marking/dropping probability towards 1, ensuring that after some number of accepted packets, the marking/dropping probability will reach 1, performing a packet marking/dropping operation. This avoids long intermarking intervals, which helps in controlling the gateway queue size effectively, and preventing congestion. From Figure 3, it is noticeable that the larger the initial marking/dropping probability, the smaller is the number of accepted packets required to direct the marking/dropping probability to 1 and hence, the less the delay before a packet mark/drop operation is activated. This is logic because a larger initial marking/dropping probability warns of the onset of congestion in near future, and therefore the uniformization process performs a packet mark/drop immediately, and thus, the sources are notified about the congestion early. In case $count.P \geq 2$, the final packet mark/drop probability is set to 1. This is logic because 2-$count.P < 0$ only happens when either the number of accepted packets (*count*) or the initial mark/drop probability $P$ or both are comparatively large values. A large value of *count* signifies that a long period of time has passed since the last packet was marked or dropped. A large value of $P$ signifies the serious deficiency of resources at the gateway caused by congestion. In both cases, it is required to perform a packet mark/drop operation immediately. The expected packet marking/dropping time examination ($E(T_m)$) is used to show how uniformization process educes clustered packet marks/drops. For a predetermined initial packet marking/dropping probability $P_{ini}$, if $P_{ini}$ is immediately applied to the arriving packets, the packet marking/dropping time ($T_m$) defined as the number of accepted packets between two successive marked/dropped packets, is geometrically distributed with $P_{ini}$ for which

$$p(T_m = n) = (1 - P_{ini})^{(n-1)} \cdot P_{ini} \qquad (7)$$

and $E(T_m) = 1/P_{ini}$ [1]. Nevertheless, if the final marking/dropping probability $P_{fin}$ is applied to the arriving packets, the packet marks/drops interval ($T_m$) will be uniformly distributed with

$$P = \begin{cases} \dfrac{P_{ini}}{2} & n \in \left[1, \dfrac{2}{P_{ini}}\right] \\ 0 & otherwise \end{cases} \qquad (8)$$

and $E(T_m) = (1/P_{ini}) + (1/2)$. Figure 4 shows the expected packet marking/dropping intervals for the geometric distribution and uniform distribution cases.

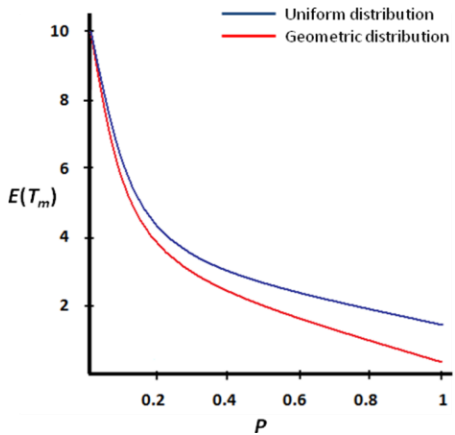

Figure 2. Expected Packet Marking/Dropping Times – Uniform Distribution: $(1/P_{ini}) + (1/2)$, Geometric Distribution: $1/P_{ini}$

From Figure 4, it is noticeable that both curves become almost parallel as $P_{ini}$ goes toward 1. Figure 4 verifies that for a predetermined marking/dropping probability $P_{ini}$, the expected packet marking/dropping time is smaller for the geometrically distributed case compared to the uniform one.

The increase in the packet marking/dropping interval is more significant for larger values of the marking/dropping probability $P_{ini}$. This indicates that the uniformization procedure increases the small expected packet marking/dropping times, as a result of large initial packet marking/dropping probabilities, ensuring that clustered packet marks/drops are minimized.

IV. CONCLUSION

This paper shows how FN uniformization process of packet intermarking intervals ensures packet drops/marks at fairly regular intervals. Avoidance of large intermarking intervals can help in controlling congestion by sending rate congestion notification signals to traffic sources in moderation on a regular basis while avoiding small intermarking intervals can help in minimizing clustered packet drops/marks and global synchronization.

AUTHORS PROFILE

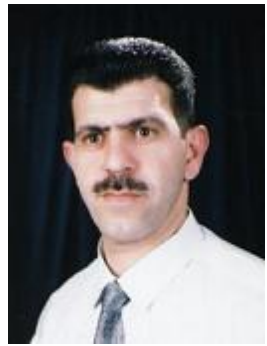

**Mohammed M. Kadhum** is a lecturer in the Graduate Department of Computer Science, Universiti Utara Malaysia (UUM) and is currently attached to the InterNetWorks Research Group at the UUM College of Arts and Sciences as a doctoral researcher. He is currently pursuing his PhD research in computer networking. His current research interest is on Internet Congestion. He has been awarded with several medals for his outstanding research projects. His professional activity includes being positioned as Technical Program Chair for International Conference on Network

Applications, Protocols and Services 2008 (NetApps2008), which has been held successfully in the Universiti Utara Malaysia. To date, he has published various papers including on well-known and influential international journals.

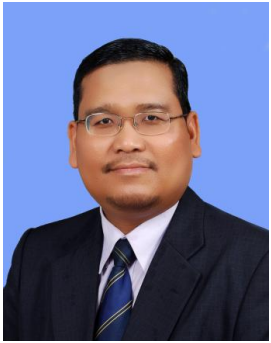

**Associate Professor Dr. Suhaidi Hassan** is currently the Assistant Vice Chancellor of the College of Arts and Sciences, Universiti Utara Malaysia (UUM). He is an associate professor in Computer Systems and Communication Networks and the former Dean of the Faculty of Information Technology, Universiti Utara Malaysia.

Dr. Suhaidi Hassan received his BSc degree in Computer Science from Binghamton University, New York (USA) and his MS degree in Information Science (concentration in Telecommunications and Networks) from the University of Pittsburgh, Pennsylvania (USA). He received his PhD degree in Computing (focussing in Networks Performance Engineering) from the University of Leeds in the United Kingdom.

In 2006, he established the ITU-UUM Asia Pacific Centre of Excellence (ASP CoE) for Rural ICT Development, a human resource development initiative of the Geneva-based International Telecommunication Union (ITU) which serves as the focal point for all rural ICT development initiatives across Asia Pacific region by providing executive training programs, knowledge repositories, R&D and consultancy activities.

Dr. Suhaidi Hassan is a senior member of the Institute of Electrical and Electronic Engineers (IEEE) in which he actively involved in both the IEEE Communications and IEEE Computer societies. He has served as the Vice Chair (2003-2007) of the IEEE Malaysia Computer Society. He also serves as a technical committee for the Malaysian Research and Educational Network (MYREN) and as a Council Member of the Cisco Malaysia Network Academy.